# Coherent photon coincidence spectroscopy of single quantum systems

Matthew Otten[1], Tristan Kenneweg[2], Matthias Hensen[3], Stephen K. Gray[1], and Walter Pfeiffer[2]

[1] *Center for Nanoscale Materials, Argonne National Laboratory, 9700 Cass Ave., Lemont, USA*

[2] *Fakultät für Physik, Universität Bielefeld, Universitätsstraße 25, 33615 Bielefeld, Germany*

[3] *Institut für Physikalische und Theoretische Chemie, Universität Würzburg, Am Hubland, 97074 Würzburg, Germany*

**Abstract**

Non-equilibrium photon correlations of coherently excited single quantum systems can reveal their internal quantum dynamics and provide spectroscopic access. Here we propose and discuss the fundamentals of a coherent photon coincidence spectroscopy based on the application of laser pulses with variable delay and the detection of an time-averaged two-photon coincidence rate. For demonstration, two simple but important cases, i.e., an exciton – biexciton in a quantum dot and two coupled quantum emitters, are investigated based on quantum dynamics simulations demonstrating that this nonlinear spectroscopy reveals information specific to the particular single quantum system.

**PhySH:** Femtosecond laser spectroscopy, Single molecule techniques, Optical spectroscopy, Coherent control, Quantum coherence & coherence measures

Coherent nonlinear spectroscopies provide detailed insights into the ultrafast dynamics of complex quantum systems [1] and have thus become versatile tools for investigating the fundamentals of light-matter interaction in, for example, light harvesting complexes [2] and semiconductor quantum dot assemblies [3]. The majority of these investigations are performed on large ensembles of quantum systems and hence details of the local dynamics in individual systems are lost. In contrast, "single-molecule"-type spectroscopies promise deeper insight since effects of inhomogeneous broadening are avoided yielding more detailed information about the coherent dynamics. A severe limitation arises from the involved nonlinearity. Hence, signals are rather weak in coherent nonlinear single-quantum system spectroscopy and thus an experimental realization still poses a severe challenge. However, the feasibility and usefulness of ultrafast coherent spectroscopy [4–6] and control [7–9] performed on single quantum systems (sQS) has been demonstrated. The demonstration of strong coupling between nanoantennas and single quantum emitters [10,11] has prepared the field for the investigation of ultrafast nanoscale quantum phenomena and new time-resolved spectroscopic methods sensitive to sQSs are needed. In particular, the investigation of the desired nanophotonic nonlinear optical effects on the single- and few-photon level requires new methodologies.

Here we propose using photon correlations, i.e., correlations in the emission of two or more photons from a sQS that is coherently driven by light pulses, as signal in nonlinear coherent spectroscopy. Using density matrix propagation we demonstrate that a) the signal reveals information specific for a particular quantum system and that b) under realistic assumptions the signal is sufficiently strong to allow for sQS spectroscopy. As depicted in Fig. 1a the relevant signal is a time-averaged quantity reflecting the photon correlations in the emitted light, i.e., the photon coincidence count rate [12]. Such coincidence rates can efficiently be measured by employing a beam splitter and two detectors. Note that the coincidence here does not arise from the emission of entangled photons, as has recently been investigated by Dorfman and Mukamel [13] for parametric down-conversion in an ensemble of quantum systems. Also it does not rely on the recently intensively investigated approach using non-classical light for nonlinear spectroscopy [14]. In the present case the emitted photons are correlated via the ongoing excitation after the first photon has been emitted and detected. Even for the rather simple case of a single two-level system a finite two-photon coincidence count rate is expected and the method can be applied. As demonstrated here the technique becomes interesting if a more complex quantum system is driven coherently and the photon coincidence rate is detected as function of the delay between driving pulses. By investigating two model systems, i.e., an exciton – biexciton system and two coupled quantum emitters, we demonstrate that the photon coincidence rate reveals properties and dynamics of the investigated system that are not accessible if just excited state populations are measured, for example, by fluorescence detection. We conclude that the detection of photon coincidences significantly increases the information content of the spectroscopic signal and opens new possibilities for conducting nonlinear coherent spectroscopy of sQSs. The paper is organized as follows: after presenting the basic measurement principle, the theoretical framework for calculating the photon coincidence rate is developed. This formalism is then applied to study two model systems and demonstrates the advantages of our proposed method, i.e., the Zeeman split exciton – biexciton in a single semiconductor quantum dot and two coupled quantum emitters.

The scheme of coherent photon coincidence spectroscopy is shown in Fig. 1a. An arbitrary sQS, here exemplified by two interacting two-level systems, is excited with a sequence of two pulses with variable interpulse delays $T$. The sQS is interacting with the environment and thus spontaneous emission leads to fluorescence that can be detected with high efficiency using conventional single molecule spectroscopy setups and photon counting detectors. Here, we assume that the fluorescence lifetime is so short that the emission dynamics cannot be directly resolved in the time domain by a fast photon detector, i.e., that it is below the typical response time of a photon counting detector of about $10^{-9}$ s. Hence, measuring the full transient second-order photon correlation signal is not possible since the timing of individual emission events in the emitted fluorescence, as it is done in photon correlation spectroscopy using high speed time-to-digital converters, is not applicable here. Still, as is demonstrated here, the photon coincidence rate detected using two photon detectors and a beam splitter contains information about the photon correlations and will be used in the following as the experimental observable. In the cases we consider here the emission of each photon requires at least the absorption of one photon from the driving light pulses. Intrinsically, the proposed method is a nonlinear spectroscopy method, just by detecting a coincidence rate that reflects the second-order photon correlations determined both by the excitation process and the internal quantum dynamics of the studied sQS.

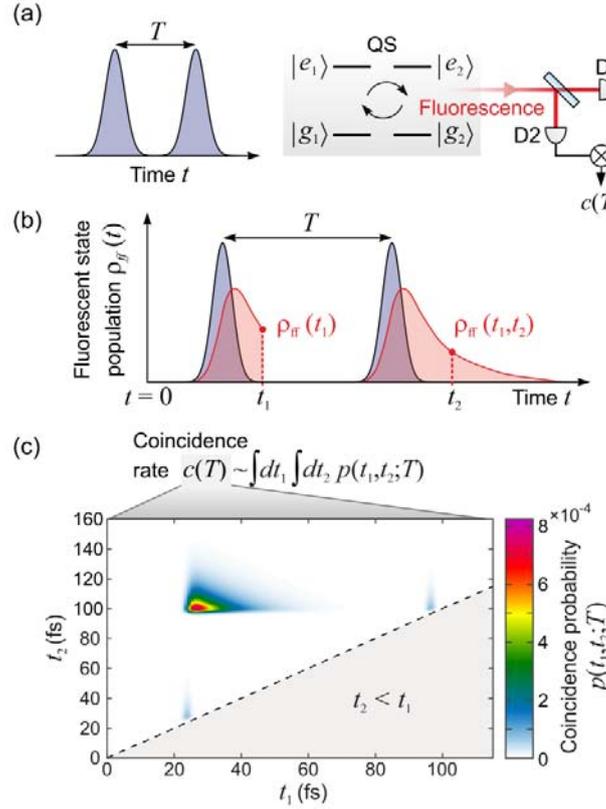

FIG. 1: Principle of coherent photon coincidence spectroscopy (CPCS). a) An arbitrary single quantum system (sQS), here exemplified by two interacting two-level systems, is periodically illuminated with a sequence of two light pulses with relative delay $T$ and the two-photon coincidence rate $c(T)$ for the emitted photons is detected using a beam splitter (BS), two photon detectors (D1, D2), and coincidence detection electronics. b) Schematic representation of the transient fluorescent state $|f\rangle$ population $\rho_{ff}(t)$ for excitation of a single two-level system with two ultrashort laser pulses (blue) and first photon emission event at $t_1$ and an indicated possible second emission event at $t_2$. Based on such population transients the photon coincidence rate is derived as is explained in the text. c) Example of a $p(t_1,t_2)$ map for a resonantly driven single two-level system characterized by the decay rate $\gamma = 3.3 \cdot 10^{-3}$ a.u. = 90 meV = $22 \cdot 10^{12}$ s$^{-1}$. Pulse delay $T = 72$ fs is chosen for two Gaussian pulses, each with 2.4 fs pulse duration.

To theoretically investigate the proposed spectroscopy method the coincidence rate has to be derived from quantum dynamical calculations. Based on the quantum regression theorem [15,16] this can be achieved without a full quantum treatment of the modes of the electromagnetic field. The probability to detect two emitted photons for excitation with a sequence of two light pulses separated by a delay $T$ is completely determined by the transient population of the fluorescent state or states in the quantum system provided that at a given earlier time $t_1$ a first photon was emitted. Formally the coincidence rate is obtained as

$$c(T) = \eta_c^2 \nu_{rep} \gamma_f^2 \int_0^{T_{rep}} dt_1' \int_{t_1}^{T_{rep}} dt_2' \, G^{(2)}(t_1', t_2'; T), \qquad (1)$$

with photon detection efficiency $\eta_c$, the excitation repetition period $T_{rep} = \nu_{rep}^{-1}$, the radiative decay rate of the fluorescent state $\gamma_f$, and the non-normalized transient two-photon correlation function $G^{(2)}(t_1,t_2;T)$ which contains information about the correlation between two photons emitted at $t_1$ and $t_2$. See Supplemental Material on page 14 in this pdf-file for a detailed derivation of Eq. (1). Based on the common definition of $G^{(2)}$ [15] (see Supplemental Material on page 14 in this pdf-file, Eq. (S1)) and the quantum regression theorem one obtains $G^{(2)}$ as

$$G^{(2)}(t_1,t_2,T) = \mathrm{tr}\left[\mathbf{a}^\dagger \mathbf{a}\, \mathbf{V}(t_2,t_1;T)\left[\mathbf{a}\,\boldsymbol{\rho}(t_1;T)\mathbf{a}^\dagger\right]\right], \qquad (2)$$

with $\mathbf{V}(t_2,t_1;T)$ as propagator from $t_1$ to $t_2$ and $\mathbf{a}$ as the operator representing the emission of a photon. The propagator $\mathbf{V}(t_1,t_2;T)$ and the density matrix $\boldsymbol{\rho}(t_1)$ at time $t_1$ of the photon emission are derived using the Liouville master equation within the Lindblad formalism [17,18] (see Supplemental Material on page 14 in this pdf-file, Eq. (S3)). To avoid treating the field modes explicitly, only the matter part of $\mathbf{a}$ which in general acts both on the matter and field modes is used here. In particular, we take $\mathbf{a}$ to be the sum of all the relevant two-level lowering operators for the given system. The numerical evaluation of Eq. (2) for given $t_1$ and $t_2$ yields the conditional probability $p(t_1,t_2;T) = G^{(2)}(t_1,t_2;T)\gamma_f^2\, dt^2$ that within two time-interval d$t$ a photon is emitted at time $t_1$ and another one is emitted at a later time $t_2$. Since the individual emission events are not resolved this quantity has to be added up for all time bins [$t_{1,2}$, $t_{1,2}$+d$t$] yielding the total probability of detecting a coincidence event per excitation cycle.

An example for a single two-level system and a given pulse delay between the two excitation pulses is shown in Fig. 1c. The contour plot shows $p(t_1,t_2;T=72$ fs$)$ and the first pulse centered at $t = 24$ fs. The coincidence probability is dominated by contributions at about $t_1 = 24$ fs and $t_2 = t_1 + T = 96$ fs, i.e., one photon is emitted during the excitation with the first pulse and the second photon emission occurs with high probability during or after the excitation with the second pulse. The wings of this maximum reflect the delayed emission of the first and second photon within the excited state lifetime of the two-level system, respectively. The small coincidence probability peaks just above the diagonal reflect the emission of two photons within either the first or the second pulse. From this probability map the coincidence rate $c(T)$ is deduced using Eq. (1) and by varying $T$ a one-dimensional CPCS spectrum can be obtained.

As a first CPCS example we discuss an exciton – biexciton system (Fig. 2a) as it may arise in a self-assembled semiconductor quantum dot, i.e., a system that has already been studied using nonlinear optical methods [19] and single quantum systems [20]. The exciton spin eigenstates $|X^+\rangle$ and $|X^-\rangle$ of a bi-axially symmetric quantum dot are degenerate for vanishing exchange interaction and can be selectively excited by circularly polarized light [21,22]. This degeneracy is lifted for example by an external magnetic field and the Hamiltonian is then given by

$$\mathbf{H} = \left(\hbar\omega_X + \delta\right)|X^+\rangle\langle X^+| + \left(\hbar\omega_X - \delta\right)|X^-\rangle\langle X^-| + 2\hbar\omega_X|XX\rangle\langle XX|, \tag{3}$$

where $\hbar\omega_X$ is the exciton energy and $\delta$ represents the Zeeman energy for the $S = \pm 1$ states of the $|X^+\rangle$ and $|X^-\rangle$ excitons. For simplicity, no biexciton binding energy is considered. The system is driven by one $\sigma^+$ polarized pulse $E^+(t)$ followed, after some delay $T$ by a $\sigma^-$ polarized pulse $E^-(t)$, both being defined as Gaussian pulses with duration $t_p$. The light-matter interaction is treated with the rotating wave approximation assuming identical dipole moments $\mu$ for $\sigma^+$ and $\sigma^-$ transitions. Identical radiative emission rates $\gamma_i$ for all Lindblad dissipation jump operators $\mathbf{J}_i \in \{|X^+\rangle\langle XX|, |X^-\rangle\langle XX|, |0\rangle\langle X^+|, |0\rangle\langle X^-|\}$ are used. The spontaneous emission processes (dashed arrows in Fig. 2a) correspond to fluorescence photon emission that are treated as indistinguishable and all photon coincidence possibilities for spontaneous emissions from $|X^+\rangle, |X^-\rangle$, and $|XX\rangle$ are added. Hence, the photon 'generation' operator $\mathbf{a}$ in Eq. (2) is given by the sum over all possible emission processes, $\mathbf{a} = \sum_i \mathbf{J}_i$.

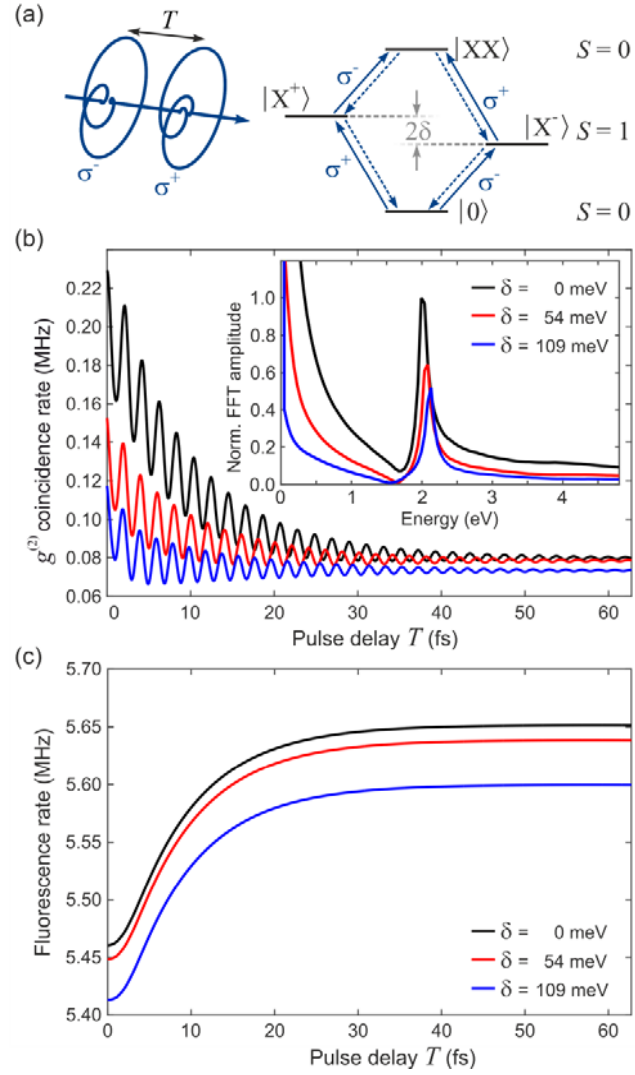

FIG. 2: Time-resolved coherent photon coincidence spectroscopy of a exciton – biexciton system ($|0\rangle$, $|X^+\rangle, |X^-\rangle, |XX\rangle$) with a splitting corresponding to the Zeeman energy $\delta$ in an external magnetic field and vanishing biexciton binding energy. a) Level scheme used for the simulations. Right ($\sigma^+$) and

left ($\sigma^-$) circular polarized pulses with variable mutual time delay $T$ are used for excitation. The excitons $|X^+\rangle$ and $|X^-\rangle$ reflect the spin $S$ eigenstate basis for excitons in a semiconductor quantum dot. The same transition matrix elements are chosen for the $\sigma^+$ and $\sigma^-$ transitions with dashed and solid arrows representing spontaneous ($\gamma$) and stimulated transitions ($\mu$), respectively. b) Two-pulse correlation coherent photon coincidence spectra calculated based on Eq. (1) in the time and, as inset, in frequency domain for different Zeeman energies $\delta$. c) Two-pulse correlation fluorescence yield spectrum calculated based on Eq. (S5). Parameters used in this simulation: $\hbar\omega_X = 7.35\cdot 10^{-2}$ a.u. = 2.0 eV; $\gamma = 3.3\cdot 10^{-3}$ a.u. = 90 meV = $22\cdot 10^{12}$ s$^{-1}$; $\mu = 3.93$ a.u. = 10 D = $33.3\cdot 10^{-30}$ Cm; $E_0 = 1.4\cdot 10^{-3}$ a.u. = $7.2\cdot 10^8$ V m$^{-1}$; $t_p = 100$ a.u. = 2.4 fs; $\eta_c = \eta_f = 0.2$; and $\nu_{rep} = 100$ MHz.

The coherent photon coincidence spectra $c(T)$ obtained for this exciton – biexciton system for different Zeeman splitting are shown in (Fig. 2b). Interestingly, in all cases the signal exhibits an interference pattern as function of $T$, whereas the corresponding fluorescence signal $f(T)$ (Fig. 2c) is not modulated. This reflects that $c(T)$ is indeed a nonlinear signal due to the detection of two photons. The fluorescence signal requires only the absorption of one photon and since the two excitation pulses are orthogonally polarized the excitations driven by both pulses are uncorrelated and lead to an unmodulated two-pulse correlation signal (Fig. 2c). In contrast, the detection of two photons requires a more complex quantum trajectory. This leads to a correlation of $\sigma^+$ and $\sigma^-$ transitions involving the biexciton state and results in the observed interferometric signal modulation. The Fourier transform of the time-resolved signal (inset of Fig. 2b) reveals an intriguing Fano-like line shape [23]. The asymmetry of the line changes with the Zeeman splitting $\delta$ and is assumed to reflect the interference of different excitation pathways in the given 4-level scheme. The two-photon correlation reflected in the coincidence signal is enhanced for $\omega > \omega_X$ whereas destructive interference of pathways dominates for $\omega > \omega_X$, indicating that the two-photon correlation in this system can be coherently controlled and, for example, the emission of a second photon can be suppressed via destructive interference. In contrast, the fluorescence signal is rather blunt and the small signal increase with $T$ just reflects the onset of exciton absorption bleaching at higher light intensities when both pulses overlap. The given example demonstrates that CPCS indeed is a nonlinear optical spectroscopy method and can reveal detailed information about a given system's dynamics.

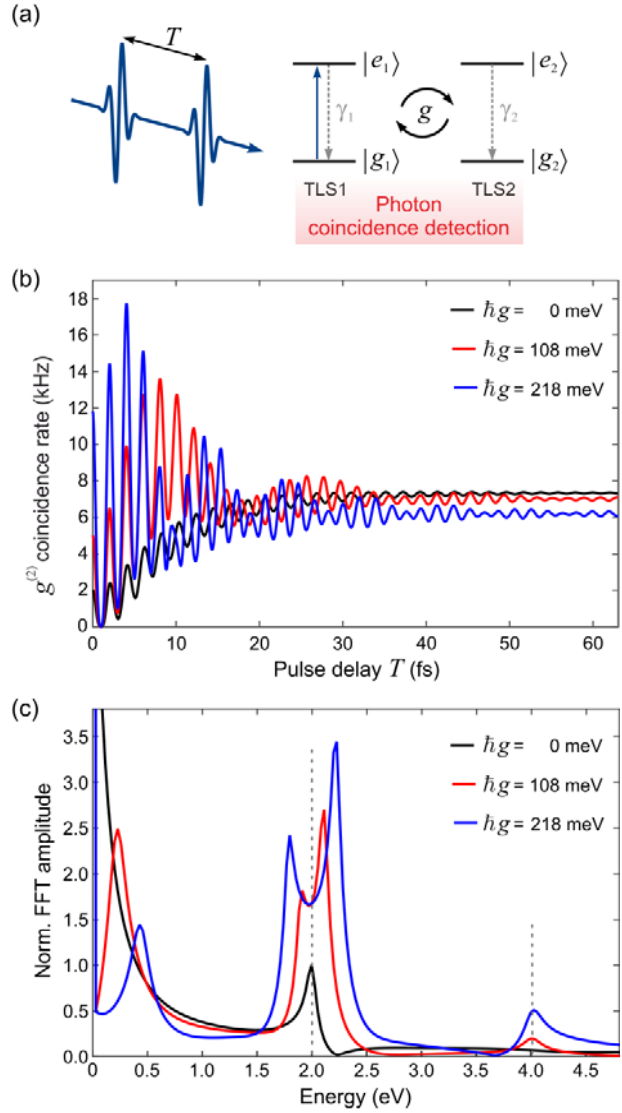

FIG. 3: Coherent photon coincidence spectroscopy for two coupled two-level systems. a) Time domain two-pulse correlation coincidence spectra for different coupling $g$ between both TLSs. Here selectively TLS 1 is excited and also fluorescence photons from TLS 1 are used for coincidence rate detection. The inset shows the excitation scheme. b) Positive axis Fourier transform magnitudes of the time-domain signals shown in part a). The vertical dashed lines indicate $\omega_0$ and $2\omega_0$, with the light pulse center

frequency $\omega_0$. Parameters used in the simulation: $\hbar\omega_0 = \hbar\omega_1 = \hbar\omega_2 = 7.35 \cdot 10^{-2}$ a.u. = 2.0 eV; $\gamma = 3.3 \cdot 10^{-3}$ a.u. = 90 meV = $22 \cdot 10^{12}$ s$^{-1}$; $\mu = 3.93$ a.u. = 10 D = $33.3 \cdot 10^{-30}$ C m; $E_0 = 2 \cdot 10^{-5}$ a.u. = $1.0 \cdot 10^7$ V m$^{-1}$; $t_p = 100$ a.u. = 2.4 fs; $\eta_c = \eta_f = 0.2$; and $\nu_{rep} = 100$ MHz.

The coincidence count rate is in the order of 100 kHz and hence is well measurable for a single quantum dot. The maximum exciton populations reached under the chosen excitation conditions are only about 0.1, i.e., a feasible excitation density for a pulsed excitation of a single semiconductor quantum dot. At these coincidence rates for a single quantum dot the here proposed coherent coincidence spectroscopy seems feasible within a reasonable data collection time of tens of seconds for a full two-pulse correlation spectrum.

As a second example, Fig. 3 summarizes CPCS of two coupled quantum emitters, each represented by a two-level system (TLS) with the Hamiltonian $\mathbf{H} = \hbar\omega_1 \boldsymbol{\sigma}_1^\dagger \boldsymbol{\sigma}_1 + \hbar\omega_2 \boldsymbol{\sigma}_2^\dagger \boldsymbol{\sigma}_2 + \hbar g \left( \boldsymbol{\sigma}_1^\dagger \boldsymbol{\sigma}_2 + \boldsymbol{\sigma}_1 \boldsymbol{\sigma}_2^\dagger \right)$, where $\hbar\omega_i$ is the excitation energy and $\boldsymbol{\sigma}_i$ the deexcitation operator for TLS $i$, and $g$ is the coupling term between both TLSs. Only TLS 1 is excited with the sequence of linearly polarized light pulses and hence $\mathbf{H}_d(t,T) = -\mu \left( E(t) + E(t-T) \right) \left( \boldsymbol{\sigma}_1^\dagger + \boldsymbol{\sigma}_1 \right)$, with $E(t)$ defined as Gaussian pulse with center frequency $\omega_0$. Photons emitted from both TLSs are collected and used for coincidence detection, i.e., $\mathbf{a} = \boldsymbol{\sigma}_1 + \boldsymbol{\sigma}_2$ in Eq. (2). An actual experimental realization of such a system might be based on the recently proposed scheme for long-distance quantum emitter coupling [24], which allows for strong coupling of two spatially separated optical nanoantennas [25].

As with the exciton – biexciton example, the case of two coupled quantum emitters yields CPC spectra that are modulated (Fig. 3) and exhibit rather complex spectral features. The overall increase of the signal reflects the TLS decay rate $\gamma$, i.e., after an initial excitation the TLS first has to decay before it can be re-excited and a second photon can be emitted. This excited state lifetime is associated with a coherence of the TLS excitation and hence for $T < 80$ fs pronounced signal modulations with $\omega_0$ appear. The saturation for large $T$ reflects the uncorrelated excitation with a second pulse and subsequent emission of another photon. For non-vanishing $g$ this saturation value drops somewhat and the signals now exhibit a coherent beating with a frequency that increases with $g$. This beating reflects the coupling induced periodic excitation exchange between both TLSs and appears in the frequency domain representation (Fig. 3b) as peak splitting by $2g$ of the $\omega_0$ peak. Interestingly, the beating signal exhibits a strong asymmetry, i.e. the signal is no longer harmonically modulated. This gives rise to a low-frequency component in the coincidence signal *directly* at $2g$. Note that no such low-frequency contribution appeared in the uncoupled exciton – biexciton system. Hence, in CPCS coupling induced state hybridization mechanisms can directly be distinguished from energy shifted eigenstates. In case of the strongest coupling, an additional component appears at $2\omega_0$. We attribute this to the very efficient energy depletion of TLS 1 via coherent excitation transfer to TLS 2 increasing thereby the nonlinearity of the coincidence signal.

Summarizing, the proposed coherent photon coincidence spectroscopy represents a new type of nonlinear spectroscopy for single quantum systems. The two simple examples discussed here served to demonstrate a) the feasibility of the method and b) that valuable dynamical information about the investigated system is revealed. In contrast to conventional nonlinear spectroscopies the method does not rely on a nonlinear signal, such as for example the generation of a second harmonic photon or of the population of a state that can only be reached in a two-step excitation process. Because of this reduced requirement the technique should be more widely applicable. Also the detection scheme relies only on a photon coincidence setup which can easily be implemented in single quantum system spectroscopy setups. Note, that the proposed technique is only a first step since the scheme can be expanded straightforwardly to a multidimensional spectroscopy by adding more excitation pulses and delays, as well as consideration of relative phases of the pulses. Higher order photon correlations such as a three-photon coincidences and a combination of different photon emission channels (polarization, wavelength, …) can be used as signals.

This work was financially supported, in part, by the German Research Foundation (DFG) within the priority program SPP1839 grant PF317-11/1 (project # 410519108). This work was performed, in part, at the Center for Nanoscale Materials, a U.S. Department of Energy Office of Science User Facility,

-12-and supported by the U.S. Department of Energy, Office of Science, under Contract No. DE-AC02-06CH11357.

and supported by the U.S. Department of Energy, Office of Science, under Contract No. DE-AC02-06CH11357.

Matthew Otten[1], Tristan Kenneweg[2], Matthias Hensen[3], Stephen K. Gray[1], and Walter Pfeiffer[2]

[1] Center for Nanoscale Materials, Argonne National Laboratory, 9700 Cass Ave., Lemont, USA

[2] Fakultät für Physik, Universität Bielefeld, Universitätsstraße 25, 33615 Bielefeld, Germany

[3] Institut für Physikalische und Theoretische Chemie, Universität Würzburg,
Am Hubland, 97074 Würzburg, Germany

**Derivation of two-photon coincidence rate $c(T)$ for a single quantum system**

The derivation of the two-photon coincidence rate $c(T)$ for a single quantum system (sQS) is based on the non-normalized transient two-photon correlation function $G^{(2)}(t_1,t_2)$ which contains information about the correlation between two photons emitted at $t_1$ and $t_2$. $G^{(2)}$ is commonly defined as [1]

$$G^{(2)}(t_1,t_2) = \langle \mathbf{a}^\dagger(t_1)\mathbf{a}^\dagger(t_2)\mathbf{a}(t_2)\mathbf{a}(t_1)\rangle, \tag{S1}$$

where $\mathbf{a}$ is the operator representing the emission of a photon. Note that $\mathbf{a}$ in its most general form is acting both on the matter system and the photon field, since the generation of a photon always requires a deexcitation in the matter system. As far as photon statistics are considered, mostly the field part of $\mathbf{a}$ alone is considered. To avoid treating the field modes explicitly, the matter part of $\mathbf{a}$ is used here to evaluate $G^{(2)}$. In particular, we take $\mathbf{a}$ to be the sum of all the relevant two-level lowering operators. To calculate $G^{(2)}$ numerically, we make use of the quantum regression theorem [1]. Consequently $G^{(2)}$ is given by

$$G^{(2)}(t_1,t_2) = \mathrm{tr}\left[\mathbf{a}^\dagger\mathbf{a}\,\mathbf{V}(t_2,t_1)\left[\mathbf{a}\,\boldsymbol{\rho}(t_1)\mathbf{a}^\dagger\right]\right], \tag{S2}$$

with $\mathbf{V}(t_2,t_1)$ as propagator from $t_1$ to $t_2$. The dynamics of the sQS is modelled using the Liouville master equation for the density matrix [2,3],

$$\dot{\boldsymbol{\rho}}(\tau) = -\frac{i}{\hbar}[\mathbf{H},\boldsymbol{\rho}] - \frac{i}{\hbar}[\mathbf{H}_d(\tau),\boldsymbol{\rho}] + \mathcal{L}(\boldsymbol{\rho}), \qquad (S3)$$

where $\boldsymbol{\rho}$ is the density matrix; $\mathbf{H}$ is the bare system Hamiltonian, including the sQSs states and possible coupling terms; $\mathbf{H}_d$ is the time-dependent driving term that parametrically depends on the delay $\tau$ between both pulses; and $\mathcal{L}$ is the Lindblad superoperator reflecting dissipation and decoherence processes. We numerically propagate the density matrix using an explicit 4$^{\text{th}}$ order Runge-Kutta method, allowing us to evaluate the quantum regression theorem for all possible combinations of photon emissions at $t_1$ and $t_2$, implemented in the open source code QuaC [4].

The quantum regression theorem result of Eq. (S2) corresponds to propagating some initial density matrix to a time $t = t_1$, forcing an 'emission' of a photon at this time, represented by the $\mathbf{a}\boldsymbol{\rho}(t_1)\mathbf{a}^\dagger$ term, propagating to a later time $t_2$, and measuring the probability of being excited at that time. This latter probability is, effectively, the likelihood of a second emission at $t_2$ given an emission at $t_1$. Note that the times $t_1$ and $t_2$ are not at all related to the pulse sequence in our spectroscopy.

At time $t_0$, before the onset of the first driving pulse, the system is prepared in the initial state. As soon as the excitation field has generated some population in the considered fluorescent state $|f\rangle$ there is a finite chance that a photon is emitted at $t_1$. The probability $p_1(t_1)$ that this occurs within a short time interval d$t$ is given by $\rho_{\text{ff}}(t_1)\gamma_{\text{f}}\,\text{d}t$, where $\rho_{\text{ff}}(t_1)$ is the diagonal element of the QS's density matrix corresponding to the state $|f\rangle$ and $\gamma_{\text{f}}$ is the fluorescence photon emission rate for this state. At $t_1$ the matter deexcitation operator for state $|f\rangle$, $\mathbf{a}_{\text{f}}$, acts on the quantum state and the resulting density matrix is given by $\mathbf{a}\boldsymbol{\rho}(t_1)\mathbf{a}^\dagger$, i.e., the inner square bracket in Eq. (S2). After this instantaneous event the further population dynamics is again determined by Eq. (S3). The probability $p_2(t_1,t_2)$ that another photon is emitted at $t_2$ is again given by $\rho_{\text{ff}}(t_1,t_2)\gamma_{\text{f}}\,\text{d}t$, where $\rho_{\text{ff}}(t_1,t_2) = \text{tr}(\mathbf{a}_{\text{f}}^\dagger \mathbf{a}_{\text{f}} \boldsymbol{\rho}(t_1,t_2))$. Hence $G^{(2)}(t_1,t_2)\gamma_{\text{f}}^2\,\text{d}t^2$ is the conditional probability $p(t_1,t_2) = p_1(t_1)\,p_2(t_1,t_2)$ that two photons are emitted, one at $t_1$ and one at $t_2$. Note that this quantity contains the full information about the transient photon correlation for the given driven quantum system and depends on the excitation conditions, i.e., in our case on the delay $T$ between two excitation pulses that is not explicitly contained in the above expressions. However, also other experimental parameters might be used.

Based on $p(t_1,t_2)$ the coincidence rate $c(T)$ is deduced. Since the individual emission events are not resolved the photon coincidence probability $p(t_1,t_2;T)$ has to be added up for all time bins $[t_{1,2}, t_{1,2}+dt]$ yielding the total probability of detecting a coincidence event per excitation cycle. Hence, $c(T)$ is given by

$$c(T) = \eta_c^2 \, \nu_{rep} \gamma^2 \int_0^{T_{rep}} dt_1' \int_{t_1}^{T_{rep}} dt_2' \, G^{(2)}(t_1',t_2';T), \tag{S4}$$

with the photon detection efficiency $\eta_c$ and the excitation repetition rate $\nu_{rep} = T_{rep}^{-1}$. $T_{rep}$ is chosen so that all internal excitations have decayed before the next excitation cycle starts. Analogously, the single fluorescence photon detection count rate $f(T)$ is calculated using

$$f(T) = \eta_f \, \nu_{rep} \gamma \int_0^{T_{rep}} dt' \, \mathrm{tr}\left[\mathbf{a}^\dagger \mathbf{a} \boldsymbol{\rho}(t';T)\right]. \tag{S5}$$

Note that the photon detection efficiencies $\eta_f$ and $\eta_c$ can differ since in the latter case a 50:50 beam splitter is required for coincidence detection that can be omitted if the single photon fluorescence signal is recorded.

For the examples discussed in the main manuscript in addition to the system Hamiltonian $\mathbf{H}$, the driving term $\mathbf{H}_d$, and the Lindblad superoperator $\mathcal{L}$ need to be defined. Here we use Gaussian pulses for driving the sQS and the field $E_i(t)$ for the $i$-th pulse is expressed using a field amplitude $E_0$, a Gaussian envelope function and a common carrier frequency $\omega_0$ as

$$E_i(t) = E_0 \exp\left[-2\ln 2 \left(\frac{t-T_i}{t_p}\right)^2\right] \cos(\omega_0(t-T_i)), \tag{S6}$$

with pulse duration $t_p$ and $T_i$ as time at which the $i$-th pulse interacts with the system.

Within the rotating wave approximation the driving term $\mathbf{H}_d$ for an individual pulse in eq. S3 is then given by

$$\mathbf{H}_d = -E_i(t) \sum_{m>n} \mu_{mn} \left( |m\rangle\langle n| + |n\rangle\langle m| \right), \tag{S7}$$

where all possible transitions in the given System and their corresponding dipole moments are $\mu_{mn}$ are considered for the transitions between energy eigenstates $|m\rangle$ and $|n\rangle$ of **H**, where the indices of the states are ascending with increasing energy.

The Lindblad superoperator $\mathcal{L}$ can be expressed as (see for example [1])

$$\mathcal{L}(\boldsymbol{\rho}) = \sum_i \gamma_i \left( \mathbf{J}_i \boldsymbol{\rho} \mathbf{J}_i^\dagger - 1/2 \left( \mathbf{J}_i^\dagger \mathbf{J}_i \boldsymbol{\rho} + \boldsymbol{\rho} \mathbf{J}_i^\dagger \mathbf{J}_i \right) \right), \tag{S8}$$

with rates $\gamma_i$ for all Lindblad jump operators $\mathbf{J}_i$. In case of the spontaneous emission process, i.e., a dissipation mechanism acting on the sQS, is given by $\mathbf{J}_{mn} = |m\rangle\langle n|$ with $m < n$ and $\mu_{mn} \neq 0$. Pure dephasing processes are not considered here, however, adding Lindblad terms based on decoherence jump operators would allow accounting for pure dephasing processes as well.